\documentstyle[aps]{revtex}
\input epsf

\def\hphi{\hat \phi}
\def\bphi{\bar \phi}
\def\Tr{{\rm Tr}}
\def\L{{\cal L}}
\def\hH{\hat H}
\def\r{{\bf r}}
\def\tG{{\tilde G}}
\def\bk{{\bf k}}
\def\ek{\varepsilon_{\bf k}} 
\def\half{\frac{1}{2}}
\def\bs{{\bar \sigma}}
\def\bpi{{\bar \pi}}
\def\btau{{\bar \tau}}

\def\tV{{\widetilde V}}

\def\tPhi{{\widetilde \Phi}}

\def\tphi{{\tilde \phi}}
\def\bphi{{\bar \phi}}
\def\cD{{\cal D}}
\def\cO{{\cal O}}
\def\fb{f_{\rm B}}

\def\teps{{\tilde \epsilon}}

\begin{document}
\draft


\title{Chiral Phase Transition at Finite Temperature \\
in the Linear Sigma Model}
\author{Heui-Seol Roh\thanks{Postdoctoral fellow of Japan Society 
for the Promotion of Science (JSPS)}
and T. Matsui}
\address{Yukawa Institute for Theoretical Physics, 
Kyoto University, Kyoto 606, Japan}
\date{November, 1996} 
\maketitle                          

\begin{abstract}
We study the chiral phase transition at finite temperature in the
linear sigma model by employing a self-consistent Hartree
approximation.  This approximation is introduced by imposing
self-consistency conditions on the effective meson mass equations
which are derived from the finite temperature one-loop effective
potential.  It is shown that in the limit of vanishing pion mass,
namely when the chiral symmetry is exact, the phase transition becomes
a weak first order accompanying a gap in the order parameter as a
function of temperature.  This is caused by the long range
fluctuations of meson fields whose effective masses become small in
the transition region.  It is shown, however, that with an explicit
chiral symmetry breaking term in the Lagrangian which generates the
realistic finite pion mass the transition is smoothed out irrespective
of the choice of coupling strength.
\end{abstract}


\section{Introduction} 
Chiral symmetry plays a vital role in low energy phenomenology of
pion-pion and pion-nucleon interactions\cite{Bernstein}. The symmetry
is manifest in the underlying QCD Lagrangian in the limit of vanishing
quark mass, but it must be broken spontaneously in the QCD vacuum.
Historically, the profound implication of this phenomenon to the very
existence of the pion was first conjectured by Nambu\cite{Nambu}, long
before the advent of QCD, in analogy to the BCS theory of
superconductivity, and it led Goldstone\cite{Goldstone} to explore
more general consequences of the broken symmetry discovering his
celebrated theorem.

There are many reasons to believe that this hidden chiral symmetry
becomes manifest at high enough temperatures.  One can take an analogy
to the spin systems in statistical mechanics, like the Ising or
Heisenberg models, where the spontaneous magnetization disappears as
the temperature of the system is raised above the critical
temperature: this is a prototype of the phenomena generally known as
the order-disorder phase transition\cite{Stanley}. Field theoretical
analog of the order-disorder phase transition at finite temperature
was first investigated by Kirzhnits and Linde\cite{Kirzhnits} in the
context of electro-weak interaction in cosmological setting, and then
a systematic method of computation was developed by Dolan and 
Jackiw\cite{Dolan} and others \cite{Weinberg}.

A great amount of works have appeared in the literature on the problem
of the chiral symmetry restoration at high temperatures and at high
baryon density.  Lee and Wick\cite{LeeWick} conjectured the possible
existence of ``abnormal state'' of dense cold nuclear matter where the
chiral symmetry is restored.  Baym and Grinstein\cite{Baym} studied
the chiral phase transition at finite temperature by applying the
approximation methods developed earlier in many-body theory.  These
early expositions of the problem used the linear $\sigma$ model
\cite{Gell-Mann} which is suited for studying the role of the quantum
fluctuation since the model is renormalizabile at least at the level 
of perturbative computation\cite{Lee}.  Non-perturbative aspects
of QCD at finite temperature have been studied more recently by the
Monte Carlo method for the Euclidean path integral formulations of QCD
on finite size lattice; such studies have shown that the chiral
transition is abrupt and closely related to the quark-liberating
deconfining transition\cite{lattice}. There are many other works based
on various phenomenological chiral models such as the four-fermion
(quark) interaction model of Nambu-Jona-Lasinio (NJL) type
\cite{Nambu,Hatsuda}; the non-linear sigma model of 
Weinberg\cite{Weinberg2,Gasser} which may be considered as an
effective theory of QCD at low energies.

Recent renewed interests in the problem are partially due to the
suggestion by Bjorken\cite{Bjorken} and others\cite{Rajagopal} that
there may be some observable consequences of the chiral symmetry
breaking transition if it occurs in the course of a highly
relativistic collision of hadrons or nuclei; a chiral condensate may
occasionally grow in the wrong direction as a fluctuation during the
rapid cooling of the dense hadronic matter (quark-gluon plasma)
produced in such a collision and this may result in a significant
fluctuation in the number ratio of charged and neutral pions.

In this paper we re-examine the chiral phase transition at finite
temperatures in terms of the linear sigma model.  Our purpose is to
describe the phase transition in static equilibrium in a simple
self-consistent approximation, which is well-known in the many-body
theory as the Hartree approximation.  We will not address more
difficult problem of describing the dynamics of the phase
transition\cite{Boyanovsky,Cooper} in the present work.

We adopt the linear sigma model since its simplicity to represent the
chiral symmetry is most suited for our purpose of investigating the
symmetry aspect of the phase transition.  The model has an advantage
of being renormalizable at least in the perturbative sense. 
The Lagrangian density is given by\cite{Itzykson}
\begin{equation}
{\cal L } = {\cal L}_s + {\cal L}_{sb}
\label{lagd}
\end{equation} 
with
\begin{eqnarray}
\label{lags}
{\cal L}_s & = & \overline{\psi}[i \gamma \cdot \partial + g(\sigma +
 i {\vec\pi}\cdot {\vec\tau} \gamma^5)]\psi +\frac{1}{2}((\partial
 \sigma)^2 + (\partial \vec \pi)^2) - \frac {m^2}{2}(\sigma^2 +
 \vec\pi^2) \nonumber \\ &-& \frac{\lambda}{4!}(\sigma^2 +
 \vec\pi^2)^2 , \\ 
{\cal L}_{sb} & = & - \epsilon \sigma
\label{laga}
\end{eqnarray}
where $\psi$, $\sigma$, and $\pi$ represent the nucleon, sigma, and
pion fields, respectively. The term ${\cal L}_{s}$ is symmetric and
invariant under an $SU(2)_L \times SU(2)_R$ chiral group.  The right
and left combination $\psi^R = \frac {1}{2} (1+ \gamma^5 ) \psi$,
$\psi^L = \frac {1}{2}(1- \gamma^5 ) \psi$ transform respectively
according to the representation $(1/2,0)$ and $(0,1/2)$ while the sets
$(\sigma, \vec \pi)$ belong to the (1/2,1/2) representation.  ${\cal
L}_{sb}$ is the symmetry breaking term.  Two Noether's currents
associated with (\ref{lags}), namely the vector current and the axial
vector current, are given by
\begin{eqnarray*}
\vec{V}_{\mu} & = &  \overline{\psi} \gamma_{\mu} \frac{\vec{\tau}}{2} \psi +
\vec{\pi} \times \partial_{\mu} \vec{\pi},  \\
\vec{A}_{\mu} & = &  \overline{\psi} \gamma_{\mu} \gamma_5
\frac{\vec{\tau}}{2} \psi + 
\sigma \partial_{\mu} \vec{\pi} - \vec{\pi} \partial_{\mu} \sigma 
\end{eqnarray*}  
respectively.  The equations of motion for the fields derived from the
Lagrangian density (\ref{lags}) give the PCAC relations
\begin{equation}
\partial_{\mu} \vec{A}^{\mu} = \epsilon \vec{\pi} .
\label{pcac}
\end{equation}

In the following investigation, we concentrate on the finite
temperature behavior of the theory at zero net baryon density or zero
chemical potential for baryonic charge and we will focus on the meson
sector of the Lagrangian (\ref{lags}) which is $O(4)$ invariant under
the rotation of the meson multiplet $(\sigma, \bpi)$.  This may be
justified at low temperature where the thermal creation of
baryon-antibaryon pairs is suppressed due to the large baryon mass.
This may not be the case, however, in the transition region where the
effective baryon mass becomes small\cite{Serot}.  Inclusion of
baryonic fluctuation is straightforward but we omit it here for the
consistency of our approximation scheme (see Appendix A). 
 
All informations concerning the equilibrium properties of the system
are contained in the thermal (imaginary-time) Green's functions
defined by the ensemble average of the ``time ordered'' products of
the quantum fields in the imaginary-time Heisenberg representation,
$\hphi (x) \equiv \hphi ( \r , \tau) = e^{\tau \hat H} \hphi ( \r)
e^{-\tau \hat H}$\cite{Fetter}:
\begin{equation}
G_{\beta}^{(m)} (x_{1}, \cdots, x_{m}) \equiv 
\langle T_{\tau} \hphi(x_{1}) \cdots \hphi(x_{m}) \rangle 
= \frac { \Tr \  [ e^{- \beta \hat H}  
T_{\tau} \hphi(x_{1}) \cdots \hphi(x_{m}) ] }
{\Tr \  e^{ - \beta \hat H } }
\end{equation} 
where $\hat H $ is the Hamiltonian of the system and $\beta = 1/ k_B
T$ is the inverse temperature. The time ordering operator $T_{\tau}$
rearranges the field operator with the smallest value of $\tau$ ($0 <
\tau < \beta$) at the right of the sequence.  The thermal Green's
function can be expressed in the Euclidean path integral
form\cite{Kapusta,Serot}:
\begin{equation}
G_{\beta}^{(m)} (x_{1}, \cdots, x_{m}) = Z_{\beta}^{-1} \int \cD [
\phi ] \phi (x_1) \cdots \phi (x_m) \exp [ \int_\beta d^4x \L (\phi,
\partial \phi) ]
\end{equation}
where $\L (\phi, \partial \phi)$ is the Lagrangian density of the
system in the Euclidean metric, obtained by the substitution,
$\partial \phi (x)/ \partial t \to i \partial \phi (x) / \partial
\tau$, and
\begin{equation}
Z_{\beta} = \Tr \  e^{ - \beta \hat H }
= \int  \cD [ \phi ]  \exp [ \int_\beta d^4x \L (\phi, \partial \phi) ] 
\end{equation}
is the partition function of the system.  Here we have introduced a
short hand notation for the Euclidean space-time integral: $\int_\beta
d^4x \equiv \int d^3 r  \int_0^{\beta} d\tau$.  In the above
expression the functional integration over the classical field
variable $\phi ( \r, \tau )$ defined in the range $ 0 < \tau < \beta $
is performed with periodic boundary condition: $\phi (\r, 0) = \phi
(\r, \beta)$.

The real-time Green's functions, which describe the propagation of an
external disturbance introduced in the system in equilibrium, may be
obtained from the imaginary-time Green's function by the analytic
continuation $ \tau \to it$\cite{Fetter}. Therefore, the thermal
Green's function also contains some dynamical properties of the system
(for weak perturbations) such as the effective mass of the excitations
which we wish to study in this paper.  The well-known advantage of
using the thermal Green's function is that it allows one to use
systematic perturbation series expansion and evaluate each term by the
method of the Feynman graphs.

In the next section we review the effective potential and its loop
expansion at finite temperature.  We will apply the method to
calculate the effective mass of the mesonic excitations at finite
temperature in section 3 and derive the Hartree approximation.  The
resultant gap equations, or the self-consistency conditions on the
mass parameters for sigma mesons and pion, contains ultraviolet
divergent integrals associated with the vacuum fluctuation and
therefore requires renormalization.  We encounter a difficulty in
imposing renormalization conditions to absorb the infinities in the
redefinitions of the mass parameter and the coupling constant as in
the perturbative renormalization scheme: the only consistent
renormalization conditions in our nonperturbative approximation scheme
give symmetric solutions.  This difficulty led us to adopt the
phenomonelogical approach to ignore the vacuum polarization effects
and to retain only thermal fluctuation in calculating the loop
integral in the gap equation.

In section 4 we examine the numerical solutions of the resultant gap
equation.  We will show that they exhibit a nontrivial hysteresis
behavior characteristic of the first order phase transition in the
exact chiral limit.  It will be shown that this hysteresis disappears
when we introduce explicit symmetry breaking term consistent with PCAC
irrespective of the choice of coupling. 

A short summary and conclusions are given in section 5.  Two
appendices contain the discussion on the inclusion of the baryon
fluctuation and the high temperature expansion of the loop integral.
Throughout the paper we use the natural unit: $c = \hbar = 1$.
  
\section{The Effective Potential at Finite Temperature}

In this section we review the effective potential formalism and
the loop expansion at finite temperature.  The finite-temperature  
effective potential for the linear $\sigma$ model is given up to
the one-loop order.

\subsection{Some general formalism}
\label{general} 

The thermal Green's functions (4) can be obtained from the
generating functional $W_{\beta} [ J ]$:
\begin{eqnarray}
e^{ W_{\beta}} = Z_\beta [ J ] 
& \equiv & \langle T_{\tau} \exp[ \int_\beta d^4x  \hphi(x) J(x)] \rangle 
\nonumber \\
& = & Z_{\beta}^{-1}\int  \cD [ \phi ] \exp [ \int_\beta d^4x 
\left( \L (\phi, \partial \phi) + \phi J \right) ] 
\label{genf} 
\end{eqnarray}
by the functional differentiation with respect to the ``external
source'' $J(x)$.  
For example, the thermal average of the field operator $\hphi (x)$ is
given by 
\begin{equation}
\langle \hphi (x) \rangle = 
\frac{\delta W_\beta [ J ]}{\delta J (x)}  \Biggr|_{J=0}
\end{equation} 
and the other thermal Green's functions can be extracted from the
following Taylor series by proper functional differentiations:
\begin{equation}
W_\beta (J) = \sum_{n=1} \frac{1}{n!}  \int_\beta d^4x_1 \cdots d^4x_m
G_{\beta}^{(m)} (x_1, \cdots, x_m) J(x_1) \cdots J(x_m).
\end{equation} 

The effective action $\Gamma_{\beta} [ \bphi ]$ is defined by the
(functional) Legendre transform of the generating functional:
\begin{equation}
\Gamma_\beta [ \bphi ] = W_\beta [ J ] - \int_\beta d^4 x \bphi(x) J(x)
\label{efp}
\end{equation}  
where $J(x)$ on the right hand side is considered to be a functional 
of $\bphi(x)$ determined by $\bphi (x) = \delta W_{\beta}[J] /\delta J (x)$.
It follows then that
\begin{equation}
\frac{\delta\Gamma_\beta [ \bphi ]} {\delta\bphi(x)} =
 - J(x)
\end{equation} 
and therefore the thermal expectation value of the field (8)
corresponds to a stationary point of the effective action, while the
``curvatures'' of the effective action at such local minimum are
related to the two-point thermal Green's function:
\begin{equation}
\frac{\delta^2\Gamma_\beta [ \bphi ]} 
{\delta\bphi(x) \delta\bphi (x)} \Biggr|_{J=0} 
= - \frac{ \delta J(x)}{\delta\bphi (y)} \Biggr|_{J=0}
= - G_{\beta}^{(2)} (x, y)^{-1} .
\end{equation}    
In the following analysis, we assume that the system is homogeneous
and therefore translationally invariant as in the vacuum.  In this case
the expectation value of the field becomes independent of $x$, and the
two-point thermal Green's function $G_{\beta}^{(2)} (x, y)$ becomes a
function only of the difference of two space-time arguments; its
Fourier transform is therefore given by
\begin{equation}
\tG_{\beta}^{(2)} (k) = \int_\beta d^4 x e^{ikx} G_{\beta}^{(2)} (x, y)
\end{equation}
where $k = (\omega_n, {\bf k})$ and $\omega_n = 2\pi n \beta^{-1} $
with integer $n$ is the discrete Matsubara frequency.  

The effective potential $V_\beta$ is defined by the effective action with 
the constant $\bphi (x) $:
\begin{equation}
 V_\beta (\phi_0) = - \Gamma_\beta [\bphi] /\Omega
\qquad {\rm for} \qquad \bphi (x)  = \phi_0
\end{equation} 
where $\Omega \equiv \int_\beta d^4 x = \beta \int d^3 r $ is the
four-volume of the Euclidean space-time. The thermal average of the
field is now determined by the stationary condition of the effective
potential:
\begin{equation}
\frac{\partial V_\beta (\phi_0)} {\partial \phi_0} = 0
\end{equation} 
Since the effective potential corresponds to a part of 
$\Gamma_{\beta} [\phi]$ which contains only $k=0$ mode of 
$\tphi_k = \int_\beta d^4 x \bphi (x) e^{i k x}$,
its second derivative with respect to $\phi_0$ gives the Fourier
component of two point Green's function evaluated at zero 
external momenta:
\begin{equation}
\frac{\partial^2 V_\beta (\phi_0)} {\partial \phi_0^2}
= \tG_{\beta}^{(2)} (0)^{-1}.
\end{equation} 
The four point vertex function $\lambda_{\beta} (k_1, k_2, k_3, k_4)$
at zero external momenta $k_1 = k_2 = k_3 = k_4 = 0$ is also 
determined by the fourth derivative of the effective potential:
\begin{equation}
\frac{\partial^4 V_\beta (\phi_0)} {\partial\phi_0^4} 
= \lambda_{\beta} (0,0,0,0).
\end{equation}

The path integral expression (\ref{genf}) of the generating functional 
$W_{\beta}[J]$, together with the definition of the effective 
action (\ref{efp}), allows one to derive a loop expansion of the 
effective potential $V_\beta$\cite{Jackiw,Dolan}:
\begin{equation}
\label{efec}
V_\beta (\phi_0 ) = V_0(\phi_0) + V_1^\beta (\phi_0) + 
\Delta V^\beta (\phi_0),
\end{equation} 
where the first term $V_0 (\phi_0)$, the tree approximation, is the
potential term in the classical Lagrangian.  This term is independent
of the temperature.  Temperature dependent one-loop term $V_1^\beta$
and higher loop corrections $\Delta V^\beta$ are obtained by the
following procedure: We decompose the original Lagrangian by shifting
the field, $\phi(x) = \phi_0 + \phi'(x)$:
\begin{equation}
{\cal L}[\phi] = {\cal L}_0' [\phi';\phi_0] +
 {\cal L}'_I [\phi';\phi_0] - V_0  + \hbox{linear terms} + \hbox{constant},
\end{equation} 
where ${\cal L}'_0 [\phi', \phi_0]$ is a bilinear form of the shifted
fields $\phi'(x)$ and the new interaction term ${\cal L}'_I$ contains
higher order products of $\phi'(x)$. The one-loop contribution $V_1$
to the effective potential is then given by
\begin{equation}
e^{-\Omega V_1 (\phi_0:\beta)} = \int  \cD [ \phi'] \exp[\int_\beta
d^4 x {\cal L}'_0(\phi';\phi_0)].
\label{1lefp}
\end{equation} 
Since this functional integral Gaussian it is easily performed:
\begin{equation}
V_1^\beta (\phi_0) = - \half  \int_\beta d^4 k \ln D (k, \phi_0) ,
\end{equation} 
where we have introduced a short hand notation for the integrals over the 
momenta and the sum over the discrete Matsubara frequencies:
$$
\int_\beta d^4 k \equiv \beta^{-1} \sum_n \frac{1}{(2\pi)^3} \int
d^3 k , 
$$ 
and the ``free'' propagator $D (k, \phi_0)$ is defined by
\begin{equation}
\int_\beta d^4 x {\cal L}'_0 [\phi', \phi_0] = 
\int_\beta d^4 k \frac{1}{2} D^{-1} (k, \phi_0) \tphi_k^2 .
\end{equation}
 
The multi-loop corrections $\Delta V$ is given by
\begin{equation}
e^{- \Omega\Delta V (\phi_0: \beta) } = \frac{ \int \cD [ \phi'] \exp
[ \int_\beta d^4x \left( \L_0'(\phi';\phi_0) + \L'_I(\phi';\phi_0)
\right) ]} { \int \cD [ \phi']\exp [ \int_\beta d^4x
\L_0'(\phi';\phi_0) ]}
\label{multiloop}
\end{equation} 
which can be interpreted diagrammatically as the sum of all single
particle irreducible connected graphs involving the vertex functions
given by the shifted interaction Lagrangian ${\cal L}'_I$ with no
external leg and internal lines are associated with the ``free''
propagator $D_0 (k:\phi_0)$.

In passing we note that the effective potential can be physically
interpreted as the Helmholtz free energy density of the system 
in the presence of ``self-magnetization'' $\phi_0$:
\begin{eqnarray}
V_\beta (\phi_0)  & = & a ( T, \phi_0 ) = g ( T, J ) - \phi_0 J, \\
g ( T, J ) & = & - \frac{ k_B T }{ V } \ln {\rm Tr} 
e^{-\beta \left[ \hH - \int d^3 r \hphi (r) J \right] } ,
\end{eqnarray}
where $g (T, J)$ corresponds to the Gibbs free energy density of the
system with the ``external field'' $J$.  From the thermodynamic
relations, it is related to the pressure $ P( T, \phi_0 )$ of the
system by
\begin{equation}
V_\beta ( \phi_0 ) = - P (T, \phi_0). 
\end{equation} 

\subsection{Effective potential for the linear $\sigma$ model}
\label{efpo} 

The leading terms of the loop expansion of the finite temperature
effective potential for the $O(N)$ $\sigma$ model was computed by Dolan
and Jackiw\cite{Dolan}. If we ignore the baryon sector in the $\sigma$
model Lagrangian, the meson sector of the Lagrangian has the $O(4)$
symmetry with respect to the rotation of the meson fields $\phi_a (x)
= (\sigma (x), \pi_1(x), \pi_2 (x), \pi_3 (x) )$ in the limit of
vanishing pion mass.  This symmetry is directly reflected in the
effective potential $V_{\beta} (\bphi_a)$ at all temperatures and the
spontaneous symmetry breaking is signaled by the appearance of the
minima of $V_{\beta} (\bphi_a)$ at non-vanishing $\bphi_a$.

The leading term of the loop expansion of the effective potential 
is given for the linear $\sigma$ model by 
\begin{equation}
V_0  (\bphi_a ) = 
\half m^2 \bphi^2 + \frac{1}{4!} \lambda \bphi^4 - \epsilon \bs .
\end{equation}
This term is independent of temperature and contains the symmetry
breaking term proportional to $\bs$.  As we shall see below this is
the only term which depends on the symmetry-breaking term in the
Lagrangian $\L_{sb}$ : all other terms due to quantum fluctuations
therefore should depend only on $\bphi^2 = \bphi_a \bphi_a$.

To compute the one-loop contribution $V_1^\beta$ ( and higher loop
corrections $\Delta V^\beta $), we shift each component of the fields
by a constant amount $\bphi_a$: $\phi_a (x) = \phi'_a (x) + \bphi_a$
and decompose the Lagrangian density:
\begin{equation}
\label{dela}
{\L}[\phi_a] = {\cal L}_0 (\phi'_a (x),\bphi_a ) + 
{\cal L}_I (\phi'_a (x),\bphi_a ) + \hbox{linear terms} + \hbox{constant},
\end{equation} 
where the shifted ``free Lagrangian'' ${\cal L}_0 (\phi'_a (x),\bphi_a
)$, which is bilinear in the shifted fields $\phi'_a$, is given by
\begin{equation}
{\cal L}_0 (\phi_a (x),\bphi_a )
= \half \left( \partial \phi_a \partial \phi_a - 
\phi_a m_{ab}^2 \phi_b \right)
\end{equation} 
with the mass square matrix given by
\begin{equation}
m_{ab}^2 = [m^2 + \frac{1}{6} \lambda \bphi^2]\delta_{ab}
 + \frac{1}{3}\lambda
 \bphi_a \bphi_b ,
\end{equation} 
where $\bphi^2 = \sum_a \bphi_a^2$ and hereafter all repeated field
subscript imply the sum if not otherwise indicated. 
The shifted interaction Lagrangian becomes
\begin{equation}
{\cal L}_I (\phi_a (x),\bphi_a ) = - \frac{1}{6}\lambda
\bphi_a \phi_a (x) \phi^2 (x) - \frac{1}{4!}\lambda\phi^4 (x).
\label{intl}
\end{equation} 
 
The shifted free Lagrangian $ \L_0$  determines the free propagator at 
finite temperature:
\begin{equation}
\int_\beta d^4 x \L_0 (\phi_a (x),\bphi_a ) =
\half \int_\beta d^4 k \phi_a (k) [ {\bf D}^{-1} (k, \bphi_a) ]_{ab}
\phi_b (k)
\end{equation}
where $ [ {\bf D}^{-1} (k, \bphi_a) ]_{ab} = k^2 \delta_{ab} +
m_{ab}^2$ with $k^2 = \omega_n^2 + \bk^2$. The diagonalization of this
matrix can be performed by rotating the fields to $\bphi'_a = O_{ab}
\bphi_b = (\bphi, 0, 0, 0)$: 
\begin{equation}
O_{aa'} [ {\bf D}^{-1} (k, \bphi_a) ]_{a'b'} O_{b'b}^{-1}
=
\left( 
\begin{array}{cccc}
k^2 + m_1^2     & 0 		& 0	   	& 0     	\\
0           	& k^2 +  m_2^2  & 0        	& 0     	\\
0              	& 0         	& k^2 + m_2^2  	& 0  		\\
0             	& 0           	& 0        	& k^2 + m_2^2     
\end{array} 
\right)
\end{equation}  
where
\begin{eqnarray}
\label{maub}
m_1^2 & = & m^2 + \half \lambda  \bphi^2  
\\
m_2^2 & = & m^2 + \frac{1}{6} \lambda \bphi^2 
\nonumber 
\end{eqnarray} 
and we find
\begin{equation}
D_{ab} (k ; \bphi_a) 
= \frac{1} {k^2 + m_1^2} \frac {\bphi_a \bphi_b}
{\bphi^2} + \frac{1}{k^2 + m_2^2} \left( 
\delta_{ab} -  \frac{\bphi_a \bphi_b}{\bphi^2} \right) .
\end{equation} 

The one-loop contribution to the effective potential is given by
\begin{eqnarray}
V_1^\beta (\bphi_a) & = & 
\half \int_\beta d^4 k \ln \det {\bf D} ( k; \bphi_a ) ] \nonumber \\ 
& = & - \half \beta^{-1} \sum_n \int \frac{ d^3 k } {(2\pi)^3} 
[ \ln (k^2 + m_1^2) + 3 \ln (k^2 + m_2^2) ]
\label{V10}
\end{eqnarray}
where we have indicated the sum over the Matsubara frequency $ k_0 = 
\omega_n = 2\pi n \beta^{-1}$ explicitly.  This sum may be performed 
by the following formulae:
\begin{eqnarray}
\sum_n \frac{x}{(2n\pi)^2 + x^2} & = & \half + \frac{1}{e^x - 1}
\label{sum1} \\
\half \sum_n \ln \left[ (2n\pi)^2 +x^2 \right] 
& = & \frac{x}{2} + \ln \left| 1 - e^{-x} \right| + c_0 
\label{sum2}
\end{eqnarray}
where the formula (\ref{sum1}) can be obtained by the method of
contour integration \cite{Fetter,Dolan} while (\ref{sum2}) is derived
from (\ref{sum1}) by integration over $x$.  (The constant of the
integral $c_0 = \half \sum_{n \ne 0} \ln (2n\pi)^2 $ is actually
infinite. ) The result is summarized as
\begin{equation}
V_1^\beta (\bphi_a)  =  V_1^0 (\bphi_a) + \tV_1^{\beta} (\bphi_a) 
\end{equation}
where
\begin{eqnarray}
V_1^0 (\bphi_a) & = &  \half \int \frac{ d^3 k }{(2\pi)^3} 
\left[ \ek (m_1) + 3 \ek (m_2) \right] \\
\tV_1^{\beta} (\bphi_a) & = &  \beta^{-1} \int \frac{ d^3 k }{(2\pi)^3}
\left[ \ln ( 1 - e^{ - \beta \ek (m_1)} )
+3 \ln (( 1 - e^{ - \beta \ek (m_2)} ) \right]  
\end{eqnarray}
with $ \ek (m) = \sqrt{ \bk^2 + m^2 } $. 
The first term $V_1^0$ is due to the vacuum fluctuation; this term is
divergent but is temperature independent so that it may be removed by
some renormalization procedure at zero temperature. The
temperature-dependent term $\tV_1^\beta$ can be rewritten by
integration by parts:
\begin{equation}
\tV_1^{\beta} (\bphi_a)  =  - \int \frac{ d^3 k }{(2\pi)^3}
\left[ \frac{\bk^2} { 3 \ek (m_1)} \fb (\ek (m_1) ) 
+3 \frac{\bk^2}{3 \ek (m_2)} 
\fb (\ek (m_2)) \right]
\end{equation}
where 
\begin{equation}
\fb (E) = \frac{1}{ e^{ \beta E} - 1 } 
\end{equation}
is the bose distribution function.
Recalling that the effective potential is equal to the pressure of
the system with inverted sign, this term may be interpreted physically
as due to the pressure exerted by the ideal gas of a boson with mass
$m_1$ and another kind of boson $m_2$ with three-fold degeneracy.

The multi-loop corrections $\Delta V^\beta$ may be computed from the
formula (\ref{multiloop}) by using the shifted Lagrangian densities
$\L'_0$ and $\L'_I$.  This term may be interpreted as the contribution
of the interaction among these quasi-particles (defined by $\L'_0$)
due to the residual shifted interaction given by $\L'_I$.  We note
that both $\L'_0$ and $\L'_I$ do not depend on the symmetry breaking
piece $\L_{sb}$ of the original Lagrangian density. It therefore
follows that $\Delta V^\beta$ should only depend on $\bphi^2$.

\section{Self-Consistent Hartree Approximation}
\label{sec34}

In this section we derive a self-consistent approximation scheme for
the computation of the thermal Green's function.  To motivate the
method we first examine the effective masses of the mesonic
excitations at finite temperature using the finite temperature
effective potential in one-loop approximation.  The result is used to
introduce a self-consistent Hartree approximation at finite
temperature.  We show a difficulty in the renormalization procedure
and this leads us to take a phenomenological approach to redefine our
Hartree self-consistency conditions.

\subsection{Temperature-dependent effective mass in one-loop 
approximation}
\label{mass} 

The effective potential $ V (\bphi_a)$ depends only on $\bphi^2$ in
the absence of symmetry breaking term $\L_{sb}$ in the Lagrangian.
When we include the symmetry breaking term $\L_{sb}$ in the
Lagrangian, the only term which breaks this symmetry is the last term
in $V_0$.  It then follows that the stationary conditions with respect
to the variations of $\bphi_a$ become degenerate except for the
$\sigma$ field:
\begin{eqnarray}
\frac{\partial V^\beta}{\partial \bs} & = & 
\left(  m^2 + \frac{1}{6} \lambda \bphi^2 + 
2 \frac{\partial V_1^\beta} {\partial (\bphi^2)} \right) \bs - \epsilon = 0 ,
\label{sigma}\\
\frac{\partial V^\beta}{\partial \bpi_i} & = & 
\left(  m^2 + \frac{1}{6} \lambda \bphi^2 + 
2 \frac{\partial V_1^\beta} {\partial (\bphi^2)} \right) \bpi_i = 0 .
\end{eqnarray}
It is clear that the potential minimum always appear at $\bpi_i = 0$.
In the following, we therefore assume $\bpi_i =0$ and consider only
the first condition (\ref{sigma}) which determines the potential
minimum at $\bs = \bs_0 = \sqrt{\bphi_0^2}$.
 
Having determined the equilibrium conditions for the condensate
amplitude, we now compute the second derivatives of the effective
potential at the potential minimum; this determines the inverse
two-point thermal Green's functions in one-loop approximation at
$k=0$, which correspond to the mass square of the excitation,
$\tG_{\beta, \sigma/\pi}^{-1} ( k=0 ) =  M_{\sigma/\pi}^2$:
\begin{eqnarray}
M_\sigma^2  & = &  
m_1^2 + \left( \frac{\partial^2 V_1^\beta} {\partial \bs^2} 
\right)_{\bs=\bs_0, \bpi=0} + \cdots 
\label{Ms}, \\
M_\pi^2 & = &
m_2^2 + \left( \frac{\partial^2 V_1^\beta} {\partial \bpi^2}
\right)_{\bs=\bs_0, \bpi=0} + \cdots
\label{Mp}
\end{eqnarray}
where the first terms came from the differentiations of the tree
effective potential $V_0$ and the explicit form of the 
contributions from the one-loop term (\ref{V10}) are given by 
\begin{eqnarray}
\left( \frac{\partial^2 V_1^\beta} {\partial \bs^2} \right)_{\bs=\bs_0, \bpi=0}
& = &
\lambda \int_\beta d^4 k 
\left( \frac{1}{k^2 + m_1^2} +  \frac{1} {k^2 + m_2^2} \right) 
\nonumber \\
& & \quad -  \lambda^2 \bs^2 \int_\beta d^4 k 
\left( \frac{1}{( k^2 + m_1^2 )^2} + \frac{1} { 3 (k^2 + m_2^2)^2 }
\right) ,   
\\
\left( \frac{\partial^2 V_1^\beta} {\partial \pi_i^2} \right)_{\bs=\bs_0, \bpi=0}
& = & 
\lambda \int_\beta d^4 k 
\left( \frac{1}{k^2 + m_1^2} +  \frac{1} {k^2 + m_2^2} \right) . 
\end{eqnarray}
Since 
\begin{equation}
\left( \frac{\partial^2 V_1^\beta} {\partial \bs \partial \bpi_i }
\right)_{\bs=\bs_0, \bpi=0} = 0
\end{equation}
there is no $\sigma$ and $\pi$ mixing even at finite temperature.

We note that since 
\begin{equation}
\frac{\partial^2 V_1^\beta} {\partial \bphi^2} =
\half \lambda \int_\beta d^4 k 
\left( \frac{1}{k^2 + m_1^2} +  \frac{1} {k^2 + m_2^2} \right)
\end{equation}
the stationary condition (\ref{sigma}) for the $\sigma$ field  
gives the following relation for the pion mass
\begin{equation}
M_{\beta, \pi}^2 \bs_0 (\beta) = \epsilon 
\end{equation}
in the one-loop approximation. In the vacuum ($\beta = \infty$) the
condensate amplitude $\bs_0$ is equal to the pion decay constant
$f_{\pi}$ defined by the transition matrix element of the axial
current $ \langle 0 |A_{\mu}^j (x)|\pi^k (q) \rangle = i 
\delta_{jk} q_\mu f_\pi e^{-iq \cdot x} $
since $A_{\mu}^i = \sigma \partial_{\mu} \pi_i - \pi_i
\partial_{\mu} \sigma$ in our model. 
Therefore we have 
\begin{equation}
m_\pi^2 f_\pi = \epsilon
\end{equation}
in the vacuum.  The above relation guarantees that if we have exact
$O(4)$ symmetry in the Lagrangian ($\epsilon = 0$),  the pion becomes 
massless when $\bs_0 \ne 0$, consistent with the Goldstone theorem.  
   
\subsection{Self-consistency conditions for the Hartree approximation}

\begin{figure}
  \epsfxsize= 4 in  \hskip 2 in \special{picture 2}
  \centerline{\epsfbox{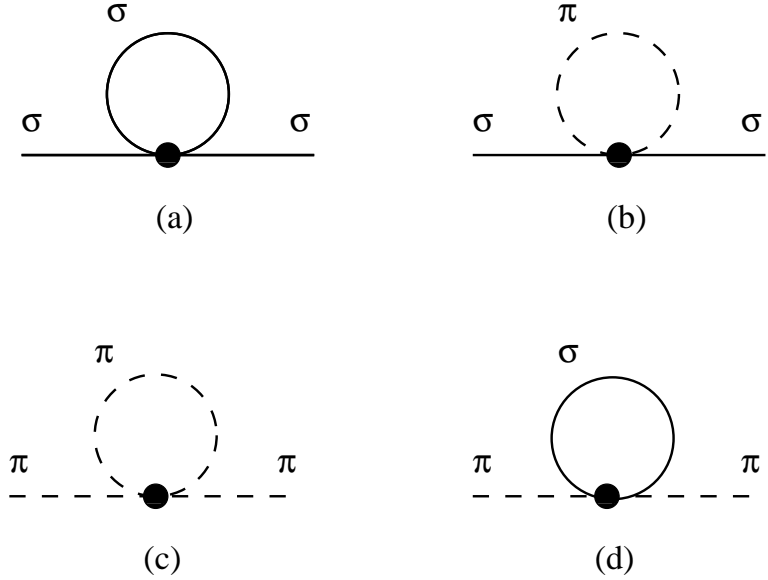}}	  
\vskip 15pt
\caption {One-loop self-energy diagrams for $\sigma$, (a) and (b), and 
for pions, (c) and (d). Solid line is the free sigma meson propagator and
the dashed line is the free pion propagator.}
\label{one-loop}
\end{figure}

The equations (\ref{Ms}) and (\ref{Mp}) are nothing but the
Schwinger-Dyson equations for the two-point thermal Green's functions
evaluated at zero momentum:
\begin{eqnarray}
\tG_{\beta, \sigma/\pi}^{-1} ( k = 0 ) =
D_{\beta, \sigma/\pi}^{-1} ( k = 0 ) + 
\Pi_{\beta, \sigma/\pi} ( k = 0 )
\end{eqnarray}
where $\Pi_{\beta, \sigma/\pi}( k )$ is the self-energy of $\sigma$ and
pions with Euclidean momentum $k = (\omega_n, \bk)$ respectively.  In
the one-loop approximation these self-energy terms are evaluated for
the one-loop Feynman graphs shown in Fig. (\ref{one-loop}) using $D_{\beta,
\sigma/\pi} ( k )$ for the internal propagators.

This observation suggests\cite{Dolan} that a partial sum of the
infinite series of the loop expansion may be made by imposing the
self-consistency condition on the one-loop result replacing the
propagators $D_{\sigma/\pi} (k)$ for the intermal lines of the
one-loop Feynmann diagrams by the Hartree propagator of the form
\begin{equation}
G_{\beta, \sigma/\pi}^{\rm H} (k) =  [ k^2 + M_{\sigma/\pi}^2 ]^{-1} .
\label{hartreeap}
\end{equation}
We {\it define} our ``Hartree approximation'' by imposing 
self-consistency conditions for the mass parameters $M_{\sigma/\pi}$
introduced in this propagator:   
masses 
\begin{eqnarray}
M_\sigma^2 & = & m^2 + \half \lambda \bphi^2 + 
\Pi_{\beta, \sigma}^{\rm H} ( M_\sigma , M_\pi ) ,
\label{msigma} \\ 
M_\pi^2 & = & m^2 + \frac{1}{6} \lambda \bphi^2 +
 \Pi_{\beta, \pi}^{\rm H} ( M_\sigma, M_\pi )
\label{mpi}
\end{eqnarray} 
where two self-energy terms are given by
\begin{equation}
\Pi_{\beta, \sigma}^{\rm H} ( M_\sigma, M_\pi ) =
\Pi_{\beta, \pi}^{\rm H} ( M_\sigma , M_\pi ) =
 \lambda \left[ \Phi_\beta (M_\sigma^2 ) + \Phi_\beta ( M_\pi^2 )
\right]
\label{mself}
\end{equation} 
with
\begin{equation} 
\Phi_\beta ( M^2 ) \equiv \int_{\beta, B} d^4 k \frac{1} { k^2 + M^2  } 
\end{equation} 
where the subscript $B$ is just to remind that the sum is taken over
the boson Matsubara frequency.  These are the generalization of the
Dolan-Jackiw ``gap equations'' for multi-component fields in the
presence of non-vanishing condensate $\bs_\beta = 0$.  This
approximation is called ``modified Hartree approximation'' by Baym and
Grinstein.  Successive substitutions of the right hand sides of the
equations into the arguments of the self-energy terms generate an
infinite series of ``superdaisy diagrams''.

\begin{figure}
  \epsfxsize= 6 in  \hskip 2 in \special{picture 2}
  \centerline{\epsfbox{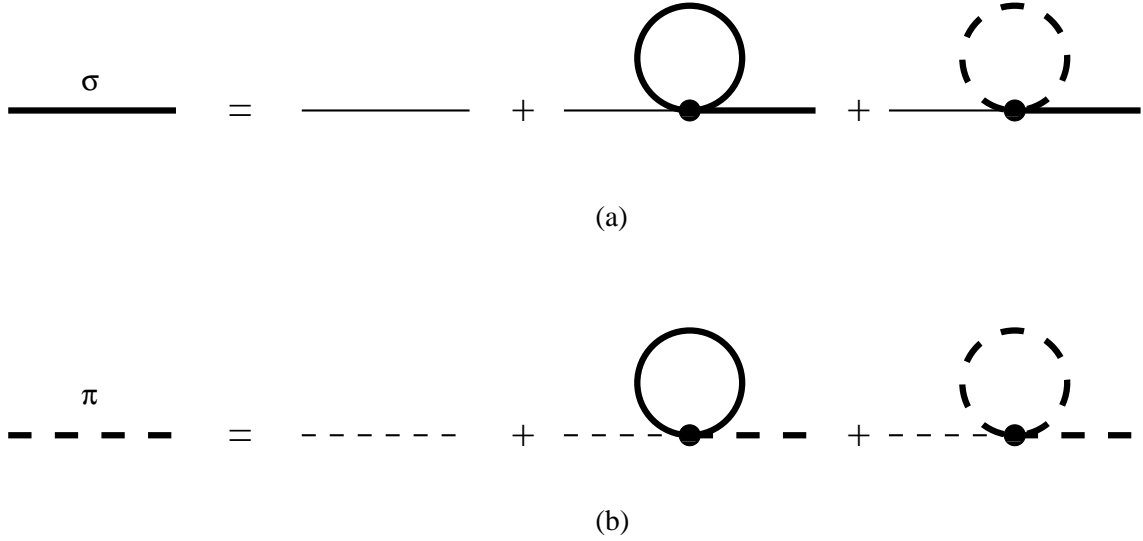}}	  
\vskip 15pt
\caption{Schwinger-Dyson equations in the self-consistent Hartree 
approximation for thermal sigma meson propagator (a)
and for thermal pion propagator (b).}
\label{hartree}
\end{figure}

We note that the direct substitution of the internal propagators in
the one-loop self-energy for the $\sigma$ field would generate an
extra term for the $\sigma$ self-energy $\Pi_{\beta, \sigma}^{\rm H}$,
\begin{equation}
\Delta \Pi_{\beta, \pi} ( k : M_\sigma , M_\pi ) =
- \lambda^2 \bs^2 \left[ \Phi_\beta' ( k: M_\sigma^2 ) 
+ \frac{1}{3} \Phi_\beta' ( k : M_\pi^2 ) \right]
\end{equation}
where 
\begin{equation}
\Phi_\beta' ( k: M^2 ) \equiv \int_B d^4 k' 
\frac{1} { (k'^2 + M^2)((k' +k)^2 + M^2 ) }.
\end{equation}
We have excluded this term since it has non-trivial $k$ dependence (or
``dispersion'') and therefore invalidates our simple ansatz
(\ref{hartreeap}) for the Hartree propagator.

\begin{figure}
  \epsfxsize= 4 in  \hskip 1 in \special{picture 2}
  \centerline{\epsfbox{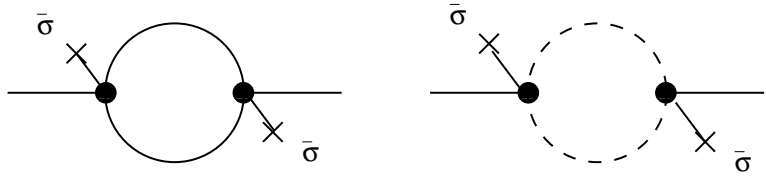}}	  
\vskip 15pt
\caption{The diagrams omitted in our approximation.}
\label{omitted}
\end{figure}

Our prescription for the self-consistent Hartree approximation is
not complete yet since we have not specified the condition for 
determining the condensate amplitude $\bs$ which appears in the
self-consistency condition for $\Pi_{\beta, \sigma}^{\rm 1-loop}$.
Our guiding principle here is the Goldstone theorem.  We replace 
the one-loop stationary condition for the $\sigma$ field by 
\begin{equation}
\left[ m^2 + \frac{1}{6} \lambda \bs^2 + 
\Pi_{\beta, \pi}^{\rm H} ( M_\sigma, M_\beta ) \right] \bs
= \epsilon .
\end{equation}
This equation, together with (\ref{mpi}), implies
\begin{equation}
M_\pi^2 \bs = \epsilon
\end{equation}
so that it guarantees that the pion becomes massless Goldstone mode in
the limit of $\epsilon \to 0$.
 
\subsection{Renormalization}
\label{ren}
The self-energy terms in one-loop approximation and in self-consistent
Hartree approximation contain (quadratic) divergences which all
originate from the divergent integral of $\Phi_\beta (M)$.  These
divergence must be removed carefully by the renormalization procedure
due to the nonlinear structure of the equations.  It turns out,
however, that such non-perturbative renormalization is possible only
when there is no symmetry breaking $\bs = 0$ as we shall see below. 

We first isolate the divergence in the self-energy terms by performing
the Matsubara frequency sum using the formula (\ref{sum1}):
We find
\begin{equation}
\Phi_\beta (M) =  \int \frac{ d^3 k}{(2\pi)^3} \frac{1}{2 \ek}
+ \tPhi_\beta (M)
\end{equation} 
where 
\begin{equation}
\tPhi_\beta (M)  =  \int \frac{ d^3 k}{(2\pi)^3} \frac{1}{\ek} 
\fb (\ek) .
\end{equation} 
All divergences in the self-energy terms arise from the phase
space integral of the first temperature independent term.
The phase space integral of $\tPhi_\beta (M)$ converges due to
the ultraviolet cut-off by the distribution function $\fb (ek)$ and
this term vanishes at zero temperature.
To regularize the divergent integral we introduce the ultraviolet
momentum cut-off $\Lambda$ in the phase space integral: 
\begin{equation}
\int_\Lambda \frac{ d^3 k}{(2\pi)^3} \frac{1}{2 \ek}  =  
I_1(\Lambda) - M^2 I_2 (\Lambda, \mu) + 
 \frac{M^2}{16 \pi^2} \ln \frac {M^2}{\mu^2} + \cO (\Lambda^{-1})
\label{regular}
\end{equation}
where
\begin{eqnarray}
I_1(\Lambda) & \equiv &  \frac{\Lambda^2}{8 \pi^2}, \\
I_2 (\Lambda, \mu) & \equiv &  
\frac{1}{16 \pi^2} \left[ \ln (4\Lambda^2/\mu^2) - 1 \right]
\end{eqnarray}
and $\mu$ is an arbitrary constant which plays the
role of a renormalization scale.  

We seek the renormalization conditions for the bare parameters $( m ,
\lambda )$ so that the resultant equations are written only in terms
of the renormalized parameters $( m_R , \lambda_R )$ and contain only 
finite terms.
For this purpose it is convenient to rewrite the equations 
(\ref{msigma}) and (\ref{mpi}) in a more symmetric form.  
We first add the two equations: 
\begin{eqnarray}
M_\sigma^2 + M_\pi^2 & = & 2 m^2 + \frac{2}{3} \lambda \bs^2 + 
2 \lambda \left[ \Phi_\beta (M_\sigma ) +
 \Phi_\beta ( M_\pi ) \right] \nonumber \\
& = & 2 m^2 + 4 \lambda I_1 (\Lambda) + \frac{2}{3} \lambda \bs^2  
- 2 \lambda I_2 (\Lambda, \mu) ( M_\sigma^2 + M_\pi^2 ) 
\nonumber \\ 
& & \quad + 2 \lambda 
\left[ \frac{M_\sigma^2}{16 \pi^2} \ln \frac {M_\sigma ^2}{\mu^2} 
+  \frac{M_\pi^2}{16 \pi^2} \ln \frac {M_\pi^2}{\mu^2} \right]
+ 2 \lambda \left[ \tPhi_\beta (M_\sigma ) + \tPhi_\beta ( M_\pi ) \right] 
\end{eqnarray}
where we have used (\ref{regular}) in deriving the second line.
This equation may be rewritten as 
\begin{equation}
M_\sigma^2 + M_\pi^2  =  - 2 m_R^2 + \frac{2}{3} \lambda_R  \bs^2 + 
2 \lambda_R \left[ \frac{M_\sigma^2}{16\pi^2}\ln \frac{M_\sigma^2} 
{\mu^2} + \frac{M_\pi^2}{16\pi^2}\ln \frac{M_\pi^2}{\mu^2}
+ \tPhi_\beta (M_\sigma ) 
+ \tPhi_\beta ( M_\pi ) \right] 
\label{renormalization}
\end{equation}
by imposing the following renormalization conditions\cite{Coleman}:
\begin{eqnarray}
- \frac{m_R^2}{\lambda_R} & = & \frac{m^2}{\lambda} +  2 I_1 (\Lambda), \\ 
\frac{1}{\lambda_R} & = & \frac{1}{\lambda} +  2 I_2( \Lambda, \mu ). 
\end{eqnarray}
On the other hand, however, the subtraction of (\ref{mpi}) from
(\ref{msigma}) gives another relation
\begin{equation}
M_\sigma^2 - M_\pi^2 = \frac{1}{3} \lambda \bs^2  
\end{equation}
which contains the bare coupling $\lambda$.  
In the limit of $\Lambda \to \infty$, the bare coupling goes
$\lambda \to  0_- $ for any finite $\lambda_R$, hence we find 
\begin{equation}
M_\sigma = M_\pi 
\end{equation} 
and in this case the renormalized self-consistency condition 
(\ref{renormalization}) becomes simply 
\begin{equation}
M^2  =  -  m_R^2 + \frac{1}{3} \lambda_R  \bs^2 + 
2 \lambda_R  \left[ \frac{M^2}{16\pi^2} \ln \frac{M^2}{\mu^2} 
+ \tPhi_\beta (M) \right] 
\end{equation}
for $M = M_\sigma = M_\pi$. Unfortunately, this renormalization scheme
is possible only when the symmetry is not broken $\bs = 0$ and
$\epsilon = 0$ and does not apply for the broken-symmetry phase. 

Similar difficulty was noted in \cite{Baym} for the finite temperature
self-consistent approximation similar to ours.

Having seen the difficulty in finding the decent renormalization
conditions for our self-consistency conditions, we take an
phenomenological approach to the problem: we replace the divergent
integral $\Phi_\beta (M^2)$ by its finite temperature-dependent piece
$\tPhi_\beta (M^2)$ simply discarding the divergent integral
associated with the ``vacuum loops''.  We therefore impose 
\begin{eqnarray}
M_\sigma^2 & = & m^2 + \half \lambda \bphi^2 + 
 \lambda \left[ \tPhi_\beta (M_\sigma^2 ) + \tPhi_\beta ( M_\pi^2 )
\right] ,
\label{gap1}\\
M_\pi^2 & = & m^2 + \frac{1}{6} \lambda \bphi^2 +
 \lambda \left[ \tPhi_\beta (M_\sigma^2 ) + \tPhi_\beta ( M_\pi^2 )
\right]
\label{gap2}
\end{eqnarray} 
for the temperature-dependent effective masses and 
\begin{equation}
\left[ m^2 + \frac{1}{6} \lambda \bphi^2 +
 \lambda \left( \tPhi_\beta (M_\sigma^2 ) + \tPhi_\beta ( M_\pi^2 ) \right)
\right] \bs = \epsilon
\label{gap3} 
\end{equation}
for the condition to determine the amplitude of the condensate.

The physical rationale for this procedure is the observation that 
each term in the self-energy term, e. g. 
\begin{equation}
\lambda \tPhi_\beta (M^2 ) 
= \lambda \int \frac{d^3 k}{(2\pi)^3} \frac{1}{2 \ek (M)} \fb 
(\ek (M)) 
\end{equation} 
is just due to the forward scattering of the meson by other mesons of
effective mass $M$ present in the system.  Since the $\sigma$ model
should be considered as an effective theory of the underlying
microscopy theory of the strong interaction, namely QCD, it would be
physically sensible to only take into account the effect of physical
excitations in the system as these equations actually do. This
approximation also corresponds to neglecting the ``many-body''
interactions among the thermally excited mesons due to the vacuum
loops.

\section{Numerical analysis of the gap equations}
In this section we present numerical solutions of the gap equations
derived in the previous section.  It will be shown first that in the
exact chiral limit the solutions exhibit hysteresis behavior
characteristic of first order phase transition.  We will show,
however, that the introduction of the finite symmetry breaking term
washes away this hysteresis for any choices of model parameters in
accordance with the physical conditions. 

In constructing numerical solutions of the gap equations (\ref{gap1}),
(\ref{gap2}) and (\ref{gap3}), we first note that from (\ref{gap1})
and (\ref{gap2}) the effective sigma mass is related to the effective
pion mass by
\begin{equation}
M_\sigma^2 = M_\pi^2 + \frac{1}{3}\lambda {\bs}^2
\label{sigmapi}
\end{equation}
while (\ref{gap2}) and (\ref{gap3}) imply that the effective pion mass
is related to the strength of the explicit symmetry breaking term in
the Lagrangian density by
\begin{eqnarray}
M_\pi^2 \bs = \epsilon .   
\label{pimass}
\end{eqnarray}
What remains to be done is to solve one of the self-consistency
condition, say for the effective sigma mass,
\begin{equation}
M_\sigma^2  =  m^2 + \half \lambda \bphi^2 + 
 \lambda \left[ \tPhi_\beta (M_\sigma^2 ) + \tPhi_\beta ( M_\pi^2 )
\right] ,
\label{sgap}
\end{equation} 
by inserting (\ref{sigmapi}) and (\ref{pimass}).

Although the physical value of $\epsilon$ is actually given in terms
of the physical pion mass and the pion decay constant by the zero
temperature relation ($\epsilon = m_{\pi}^2 f_{\pi}$), we consider
$\epsilon$ as an external variable and examine how the solutions of
the gap equations depend on the strength of $\epsilon$. 

\subsection{Exact chiral limit ($\epsilon = 0$):} 
We first examine the special case when $\epsilon = 0$.  In this
limiting case the chiral symmetry is the exact symmetry of the
Lagrangian and according to the Goldstone theorem the pion becomes
massless in the low temperature phase where the symmetry is
spontaniously broken.  Our approximation scheme in fact guarantees
this condition by (\ref{pimass}).  At zero temperature, the above 
relations (\ref{sigmapi}) and (\ref{sgap}) imply
\begin{eqnarray}
M_{\sigma}^2 ( T = 0 ) & \equiv &  m_0^2 = -2m^2 , 
\label{m0} \\ 
\bs^2 ( T = 0 ) & \equiv & \phi_0^2 = -6m^2/\lambda. 
\label{phi0}
\end{eqnarray}

In the symmetry broken phase where $\bs \ne 0$, $M_{\pi}=0$ and
$M_{\sigma}^2 = \frac{1}{3} \lambda \bs^2$, the self-consistency
condition for the effective sigma meson mass becomes
\begin{equation}
{\tilde M}_{\sigma}^2 = {\tilde m}^2 + \frac{3}{2} {\tilde M}_{\sigma}^2 
+ \frac{\lambda}{2\pi^2} \left[ 
I^{(2)}_- ( {\tilde M}_{\sigma} ) + I^{(2)}_- ( 0 ) \right]
\label{ltgap}
\end{equation}
where we have introduced dimensionless variable ${\tilde M}_\sigma
\equiv M_\sigma \beta$, ${\tilde m} = m \beta$ and 
the dimensionless function $I^{(2)} (\mu)$ is defined by
\begin{equation}
I_{\pm}^{(2)} (\mu) \equiv  \int_0^{\infty} \frac{x^2 dx}
{\sqrt{x^2 + \mu^2}} \frac{1}{ e^{\sqrt{x^2 + \mu^2}} \pm 1 } 
\end{equation}
The high temperature expansion of $I_{\pm}^{(2)} (\mu)$ appropriate
for small value of $\mu$ is given in the Appendix B.  On the other hand,
in the high temperature symmetric phase, where ${\tilde \sigma} =0$
and $M_\sigma = M_\pi$, the gap equations become degenerate to
\begin{equation}
{\tilde M}_{\sigma}^2 = {\tilde m}^2 + 2 \lambda \frac{1}{2\pi^2} 
I^{(2)}_- ( {\tilde M}_{\sigma} ) 
\label{htgap}
\end{equation}

Using $I_{\pm}^{(2)} (0) = \pi^2 /6$ 
the above two conditions are further simplified to
\begin{equation}
\frac{6}{\pi^2} I^{(2)}_- ( {\tilde M}_{\sigma} ) = 
\frac{12}{\lambda} \left[ - {\tilde m}^2 
- \frac{1}{2} {\tilde M}_{\sigma}^2 \right] - 1 
\label{lt}
\end{equation}
for the low temperature phase, and 
\begin{equation}
\frac{6}{\pi^2} I^{(2)}_- ( {\tilde M}_{\sigma} ) = 
\frac{6}{\lambda} \left[ - {\tilde m}^2 + {\tilde M}_{\sigma}^2 \right] 
\label{ht}
\end{equation}
for the high temperature symmetric phase. 

\begin{figure}
  \epsfxsize= 6 in  \hskip 2 in \special{picture 2}
  \centerline{\epsfbox{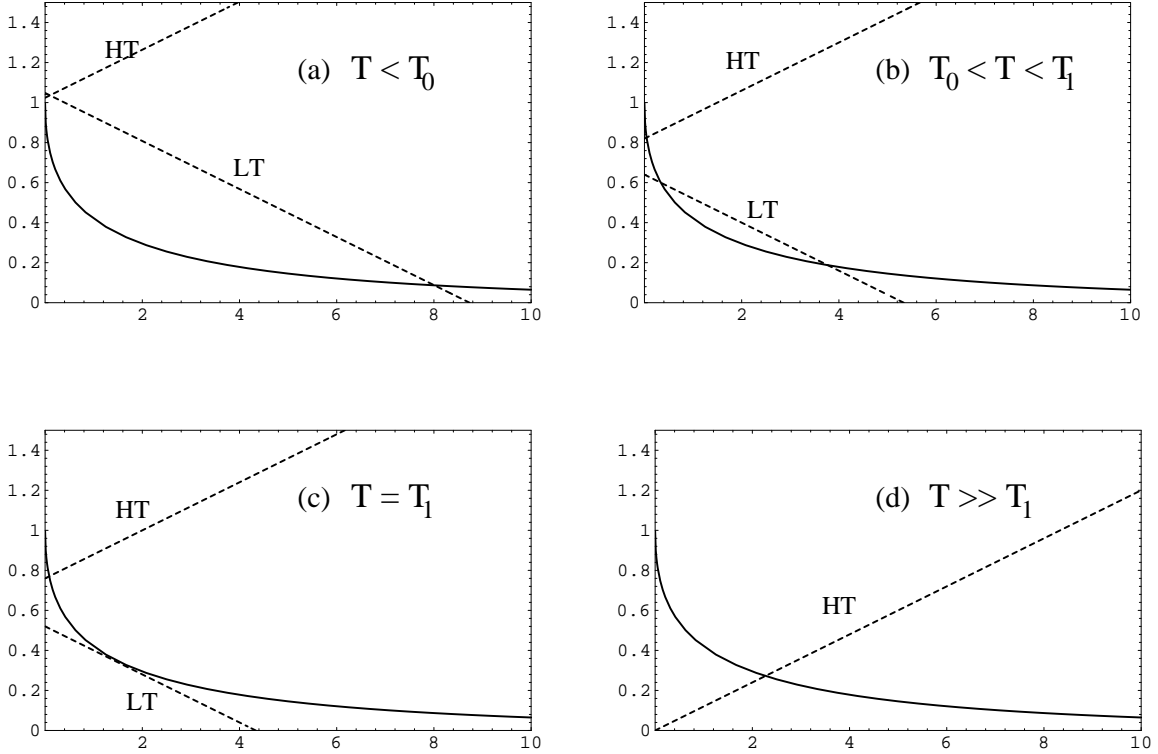}}	  
\vskip 15pt
\caption {Graphical construction of the solutions of the gap equations
(\ref{lt}) and (\ref{ht}): the solid curve depicts the left hand side
of the equations and two dashed lines marked by LT and HT are for the
right hand side of (\ref{lt}) and (\ref{ht}), respectively, both as a
function of ${{\tilde M}_\sigma}^2$.  Crossing points of the solid
curve and the dashed line give the solution of the corresponding gap
equation. }
\label{gapfig1}
\end{figure}

It is instructive to examine the solutions of the above equations
graphically.  In Fig. \ref{gapfig1} we plot the function 
\begin{equation}
f(x^2) = \frac{6}{\pi^2} I^{(2)}_- ( x ),
\end{equation}
which appear on the left hand side of the above two equations, by the
solid line; the right hand sides of (\ref{lt}) and (\ref{ht}) are
plotted by dashed curves. The intersection of two curves determines
the solutions of the scaled gap equation, (\ref{lt}) or (\ref{ht}),
for the scaled sigma mass square ${\tilde M}_\sigma^2$.  We note that
$f(z)$ is a concave, monotonically decreasing function of $z={\tilde
M}_{\sigma}^2$ normalized as $f(0)=1$, while the right hand sides of
(\ref{lt}) and (\ref{ht}) are both linear in $z$.  Since the value of
$-m^2$ is fixed by the amplitude of vacuum condensate at $- m^2 = 1/6
\lambda \phi_0^2 ( > 0 )$, the $y$-intercepts of the dashed curves 
($- 12{\tilde m}^2 /\lambda - 1$ for the low temperature phase; $- 6
{\tilde m}^2 /\lambda$ for the high temperature phase) both move
upward as the temperature is lowered, while the solid curve remains
unchanged.

In the high temperature phase, two curves cross when $- 6{\tilde
m}^2/\lambda > 1$.  The condition $- 6{\tilde m}^2/\lambda = 1$
determines the ``critical'' temperature 
\begin{equation}
T_0 = \phi_0
\label{tc} 
\end{equation} 
below which the real solution of (\ref{ht}) does not exist. Note that
$T_0$ does not depend on the strength of the interaction $\lambda$.
When one approaches to $T_0$ from high temperature side, the effective
sigma mass decreases continuously and vanishes at $T = T_0$.  On the
other hand, at low temperatures below $T_0$, where $- 6{\tilde
m}^2/\lambda > 1$, the dashed curve for the low temperature phase
intersects with the solid one at one point giving a unique solution to
the original gap equations.  As the temperature increases, the
effective sigma mass again decreases.  However, it does not vanish at
$T_0$ in this case.  Instead, as the temperature increases slightly
above $T_0$, there appears another crossing point at small ${\tilde
M}_\sigma^2$ due to the concave shape of the function $I^{(2)}_- (
\mu)$ as plotted against $\mu^2$.  As the temperature is further
increased, the two crossing points approach toward each other and they
eventually annihilate at certain temperature $T_1 ( > T_0)$ when the
dashed line becomes tangent to the solid curve.

\begin{figure}
  \epsfxsize= 6 in  \hskip 2 in \special{picture 2}
  \centerline{\epsfbox{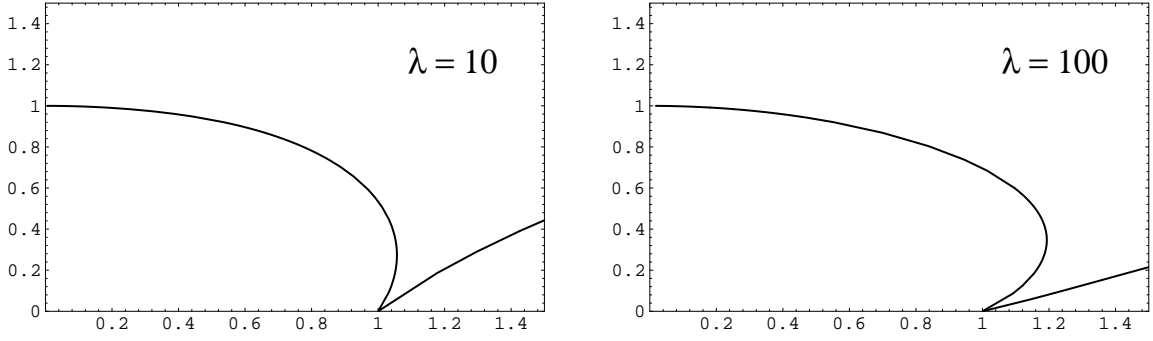}}	  
\vskip 15pt
\caption {Temperature dependence of the solutions of the gap equations 
in the chiral limit ($\epsilon = 0$) for the scaled sigma mass
($\mu_\sigma = M_\sigma/m_0$) and the scaled sigma condensate
($\chi=\bs/\phi_0$) as a function of $t = T/T_0$. In the high
temperature phase, the effective pion mass becomes degenerate with the
effective sigma mass while it vanishes in the low temperature symmetry
broken phase.  The scale of the temperature is set by $T_0 =
\phi_0$.}
\label{gapfig2} 
\end{figure}

We show in Fig. \ref{gapfig2} the temperature dependence of the
effective sigma mass and the amplitude of the sigma condensate,
normalized by their zero temperature values given by (\ref{m0}) and
(\ref{phi0}) respectively.  This plot is made by solving (\ref{lt})
and (\ref{ht}) respectively for $-{\tilde m}^2$ and making parametric
plot of the temperature ($\propto 1/\sqrt{-{\tilde m}^2}$) vs the
scaled sigma mass ($\propto {\tilde M}_\sigma / \sqrt{-{\tilde
m}^2})$.  The scale of the temperature in this plot is set by the
value of $T_0$ which is equivalent to the amplitude of the vacuum
condensate by (\ref{tc}).  The system exhibits a typical hysteresis
behavior of first order phase transition in the exact chiral limit.
We note that this hysteresis appears always irrespective to the value
of $\lambda$ since the slope of the function $f(z)$ diverges at $z=0$. 
The magnitude of $T_1$ depends on $\lambda$ explicitly, however.  For
larger $\lambda$, the slope of the dashed lines in Fig. \ref{gapfig1}
(b) becomes less steeper, thus $T_1$ is larger.  Since $T_0$ is
independent of $\lambda$, this implies that the region of the
hysteresis is larger for stronger coupling.  We next examine how this
behavior is modified when $\epsilon$ is finite.

\subsection{For $\epsilon \ne 0$:}
As we have noted earlier, $\epsilon$ plays the role of externally
applied magnetic field in the case of magnetic phase transition,
therefore we expect that the transition will get smoother for
non-vanishing $\epsilon$ .

For our numerical analysis with non-vanishing $\epsilon$, it is more
convenient to use the following dimensionless variables:
\begin{eqnarray}
\mu_\sigma & \equiv & M_{\sigma} / m_0,  \\
\mu_\pi & \equiv & M_{\pi} / m_0, \\
\chi & \equiv & \bs / \phi_0, \\
t & \equiv & T / T_0 = T / \phi_0,  \\
{\teps} & \equiv & \epsilon / ( m_0^2 \phi_0 ), 
\end{eqnarray} 
where $m_0$ and $\phi_0$ are the sigma mass and the sigma condensate
at zero temperature in the limit $\epsilon = 0$ as determined by 
(\ref{m0}) and (\ref{phi0}), respectively.
The equations (\ref{sigmapi}), (\ref{pimass}) and (\ref{sgap}) then
reduce to 
\begin{eqnarray} 
\mu_\sigma^2 & = &\mu_\pi^2 + \chi^2  ,
\label{mus}\\
\mu_\pi^2 \chi & = & {\teps} ,
\label{mup}\\
2 \mu_\sigma^2 & = & -1 + 3 \chi^2 + \frac{3}{\pi^2}t^2
\left[ I^{(2)}_- ( \sqrt{\lambda/3} \mu_\sigma / t ) 
+ I^{(2)}_- ( \sqrt{\lambda/3}\mu_\pi / t ) \right] ,
\label{mugap}
\end{eqnarray}
respectively.  Note that in this form the $\lambda$ dependence of the
gap equation is absorbed into the $\lambda$ dependence of the argument
of the dimensionless function $I^{(2)}_-$.  The solutions of these
coupled nonlinear equations depends on two dimensionless parameters,
$\lambda$ and $\teps$. In the limit ${\teps} \to 0$ they coincide with
the previous results.  

\begin{figure}
  \epsfxsize= 6 in  \hskip 2 in \special{picture 2}
  \centerline{\epsfbox{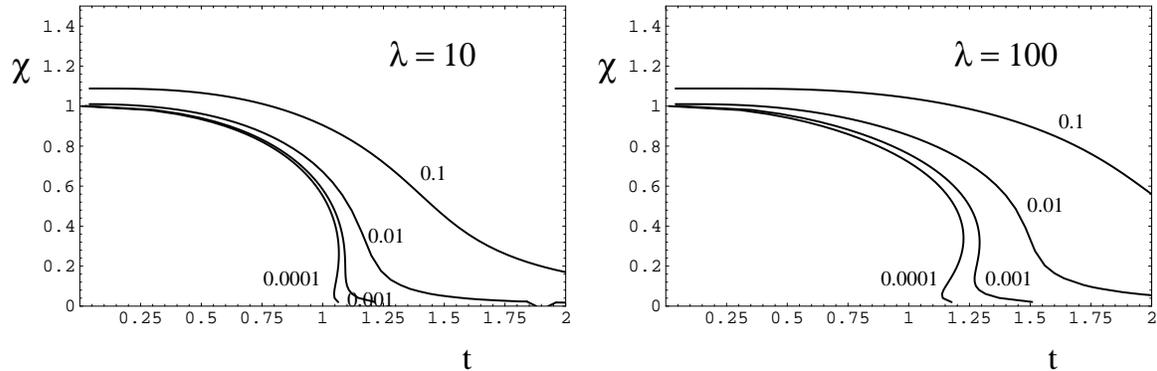}}	  
\vskip 15pt
\caption {Temperature dependence of the scaled order parameter 
$\chi = \sigma/\sigma_0$. The curves are labeled by the value of
$\teps$.  Two figures correspond to two different choices of the
coupling strength $\lambda$.}
\label{phi.fig}
\end{figure}

We show in Fig. \ref{phi.fig} the $\teps$-dependence of the scaled
order parameter $\chi$ plotted as a function of the scaled temperature
$t$. It is seen that for small values of $\teps$ the solutions exhibit
hysteresis behavior (the back-bending shape of $\chi$).  As $\teps$
increases, however, the curve stretches out gradually and for a large
value of $\teps$, $\chi $ becomes a monotonically decreasing function
of temperature.  From an inspection of the equations (\ref{mus}),
(\ref{mup}) and (\ref{mugap}), we can see that the effect of $\teps$
is expected to become significant when $\chi^2 \sim \mu_\pi^2 = \teps
/\chi$ in (\ref{mus}) or $\sqrt{\lambda/3}\mu_\pi / t \sim 1$ for the
argument of the second $I^{(2)}_-$ in (\ref{mugap}); the former
condition gives $\teps \sim \chi^3$ and the latter $\teps \sim 3 t^2
\chi /\lambda$ for the conditions that a significant modification is
caused by $\teps$.  This explains why for a relatively small value of
$\teps$, say $\teps = 0.001$, we see in the plot that a considerable
modification appears at small $\chi$, $\chi < 0.1$, and that for
larger $\lambda$ the effect is more significant.  We found that the
critical value of $\teps$ for the disappearance of the trace of
hysteresis is $\teps_c = 0.002$ for $\lambda = 100$.

\begin{figure}
  \epsfxsize= 6 in  \hskip 2 in \special{picture 2}
  \centerline{\epsfbox{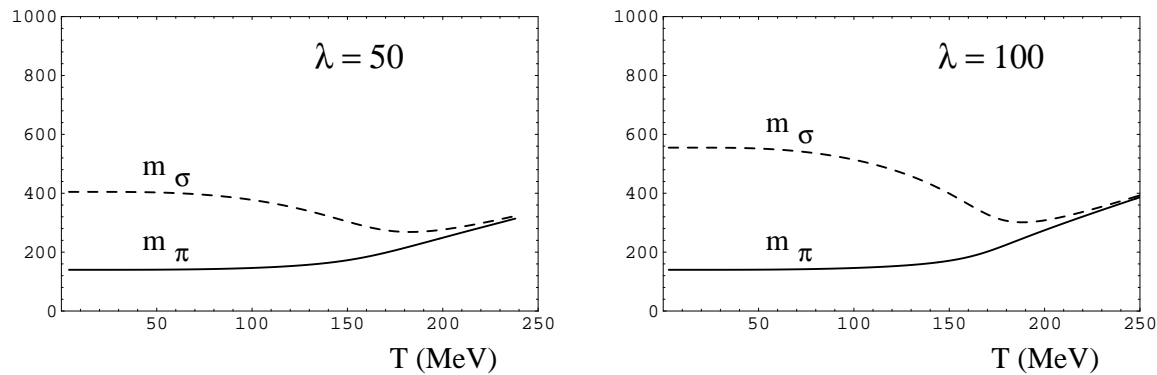}}	  
\vskip 15pt
\caption {Temperature dependence of the effective sigma mass
($M_\sigma$) and the effective pion mass ($M_\pi$) for $\epsilon$ ($ =
m_\pi^2 f_\pi$).  Two figures correspond to two different choices of
the coupling strength $\lambda$.}
\label{mass.fig} 
\end{figure}

The physical value of $\teps$ may be determined by adjusting
the values of our model parameters ($\lambda$, $-m^2$ and $\epsilon$)
so that they reproduce the physical pion mass ($m_\pi = 140$MeV), and
the pion decay constant $f_\pi = 93$MeV and the sigma mass $m_\sigma$
in the vacuum.  In the linear sigma model $f_\pi$ is identified as the
amplitude of the vacuum condensate: $\sigma_0 = f_\pi$.  Then
(\ref{sigmapi}) and (\ref{sgap}) give constraints:
\begin{equation}
m_\sigma^2  =  m_\pi^2 + \frac{1}{3}\lambda f_\pi^2 
 = m^2 + \frac{1}{2}\lambda f_\pi^2, 
\end{equation}
which implies 
\begin{equation}
\lambda = 3(m_\sigma^2 - m_\pi^2)/f_\pi^2
\end{equation}
and
\begin{equation}
-m^2 = (m_\sigma^2 - 3 m_\pi^2)/2.
\label{m2}
\end{equation}
We note that since $-m^2 > 0$ these two relations give a constraint
$\lambda > 6 m_\pi^2 / f_\pi^2 = 13.6$ on the allowed values for
$\lambda$.  When we vary $\lambda$ from 50 to 500, $m_\sigma$ varies
from 400 MeV to 1200 MeV constrained by these relations.  The
corresponding values of $\teps = \epsilon / m_0^2 \phi_0$ are
determined by using $m_0^2 = -2m^2$ and $\phi_0 = f_\pi / \chi_0$
where $\chi_0$ is the solution of the scaled gap equation at zero
temperature. Some representative results are tabulated in Table
\ref{table1}.  For large coupling $\lambda$, the sigma mass increases
as $m_\sigma \sim m_0 \sim \sqrt{\lambda/3}f_\pi$ and the relative
strength of the symmetry breaking term decreases as ${\teps} =
(m_\pi/m_0)^2 (f_\pi/\phi_0) \sim 3 \lambda^{-1} (m_\pi / f_\pi)^2$.
We show in Fig. \ref{mass.fig} the results for the temperature
dependence of the mass parameter for two different choices of the
coupling strength $\lambda$.

An interesting question is whether the hysterisis can survive for a
sufficiently large value of $\lambda$ since $\teps$ decreases for
large $\lambda$.  This is not the case, however, since the critical
value of the $\teps$ also decreases for small $\lambda$.  One can see
this point noting that the argument of the second $I^{(2)}_-$ in
(\ref{mugap}) becomes independent of $\lambda$ at large $\lambda$:
$\sqrt{\lambda/3}\mu_\pi / t \sim (m_{\pi}/f_{\pi})/\sqrt{\chi}t $ as
$\lambda \rightarrow \infty$.  For any value of $\lambda$ the introduction of
the symmetry breaking term $\epsilon = m_\pi^2 f_\pi$ with physical
value for $m_\pi$ and $f_\pi$ washes away the hysteresis behavior in
the solution of the gap equatioon.

\section{Summary}
In this paper we have studied the chiral phase transition at finite
temperature using the meson sector of the linear sigma model.  We
formulated a self-consistent approximation scheme starting from the
one-loop effective potential by imposing self-consistency conditions
on the effective meson mass equation, keeping the meson self-energy
independent of meson momentum.  The resultant equations are just the
Dolan-Jackiw gap equations for multi-component fields in the presence
of non-vanishing meson condensate.  In this approximation, the meson
mass parameter in the thermal (imaginary time) meson propagator may be
identified by analytic continuation as the mass of the real mesonic
excitations introduced in the system in equilibrium although in more
general cases such simple identification is not possible.  This
non-perturbative approximation scheme has, however, a difficulty in
choosing proper renormalization conditions to eliminate the
divergent loop integrals.  We therefore adopted a phenomenological
approach which just ignores the divergent vacuum loops.  This
procedure is technically equivalent to introduce a zero momentum
cut-off in the vacuum loop integral.

We showed that the solutions of the resultant gap equations does not
reproduce usual second order phase transition.  We found, instead,
that they exhibit hysteresis behavior characteristic of the first
order transition in the chiral limit.  It is shown that this
``premature transition'' is caused by the long range fluctuation of
the mesons fields whose effective masses become small in the
transition region.  We are however not able to calculate the
transition temperature from the gap equations alone; to do so we need
the information of the effective potential associated with the
approximation we employed in this work.  Such calculation may be
performed by the method of the composite operator effective potential
developped by Cornwall, Jackiw, and Tomboulis\cite{Cornwall}. We have
made a preliminary investigation in this direction\cite{Roh}. The
result will be reported elsewhere.

Inclusion of the symmetry breaking term generally tends to smooth
sharp phase transition, but in general, the first order transition may
survive if the symmetry breaking scale is small enough.  We found,
however, that this is not the case at least in our calculation: the
physical pion mass, together with the PCAC relation for the axial
current, gives a very strong constraint on the choice of the
parameters of our model and the hysteresis behavior is smoothed out
irrespective of the coupling strength.  We found that smooth chiral
transition takes place at $T = 150 \sim 200$MeV in our model,
consistent with recent results from the state-of-art lattice QCD
calculations at finite temperature\cite{lattice}.  An advantage of the
present approach based on the effective degrees of freedom, expressing
crucial aspects of the symmetry behavior of the system, is that it is
more tractable for investigating more difficult problem of the
dynamics of the chiral transition in high energy nuclear collisions
where we expect non-equilibrium aspects may play essential role
\cite{Rajagopal,Boyanovsky,Cooper}.  We also plan to investigate this 
problem in the future.

\section*{Acknowledgements}
We thank Brian Serot and our ex-colleagues at the Nuclear Theory
Center of Indiana University, where the present work was initiated,
for various helpful discussions at the early stage of the work.  One
of the authors (H.-S. R.) is grateful for the generous support
provided by the Japan Society for the Promotion of Science.  This work
has been supported in part by the Grant-in-Aid for Scientific Research
\#06640394 of Ministry of Education, Science and Culture in Japan.

\appendix

\section{Inclusion of Baryon}
The original $\sigma$ model Lagrangian contains the baryon (nucleon)
fields coupled to meson fields by Yukawa coupling.  At low
temperatures ($T << m_N$), we may expect the contribution of the
thermal baryon-antibaryon excitations is negligible.  This may not be
so however near the transition temperature since the baryon mass
becomes smaller due to the reduction of the $\sigma$ condensate which
dynamically generates the baryon mass.  For example, in the lowest
order of the Yukawa coupling $g$,
\begin{equation}
m_N = g \bs ,
\label{bmass}
\end{equation}
(see Fig. {baryon-self}).  We show in this appendix that the inclusion
of the baryon fluction in this approximation leads to the results very
similar to the Hartree approximation without baryons.  However, we
show also that it is not a consistent approximation for the meson
propagators.

\begin{figure}
  \epsfxsize= 1.5 in  \hskip 2 in \special{picture 2}
  \centerline{\epsfbox{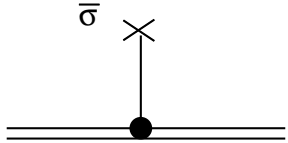}}	  
\vskip 15pt
\caption {The lowest order baryon self-energy diagram}
\label{baryon-self}
\end{figure}

Inclusion of the baryonic excitations in the computation of the
effective potential may be performed in the line similar to mesonic
excitations (Sect. 2) with one modification that the functional
integral over the Grassmann variables, which represent ``classical
fermion field'', leads to the change of the sign and the removal of
the factor $1/2$ in the one-loop contribution:
\begin{equation}
V_{1, \rm baryon}^{\beta} ( \bphi_a )  =   \int d^4 k \ln \det G_N (k)
\label{b1l}
\end{equation} 
where 
\begin{equation}
G_N (k) = \frac{1}{\not k + M_N} 
\label{baryon}
\end{equation}
is the thermal Green's function for baryons with  
$\not k \equiv i \omega_n \gamma_0 + \gamma_i k_i$ 
and the sum is taken over the fermion Matsubara frequency
$\omega_n = (2n+1)\pi \beta^{-1}$ which arises from the 
anti-periodic boundary conditions for fermionic fields 
in the path integral.  
The nucleon mass matrix is given here by
\begin{equation}
M_N = g ( \bs + i \gamma_5 {\bf \bpi}_i \btau_i ) 
\end{equation}
for non-vanishing $\bphi_a = ( \bs, {\bf \bar \pi})$.
Inserting (\ref{baryon}) into
(\ref{b1l}) and performing the determinant over Lorentz and isospin
indices we find
\begin{eqnarray}
V_{1, \rm baryon}^{\beta} ( \bphi_a ) = 
- 2 \cdot 2 \int d^4 k \ln [ k^2 + g^2 \bphi^2 ]
\end{eqnarray}
where $k^2 = \omega_n^2 + \bk^2$.  In this result, one factor 2 has arisen
from determinant in the Lorentz indices and may be attributed to the
spin degrees of freedom, while another factor 2 is due to the isospin
degrees of freedom.  Note that fermionic loop contribution is larger
than that of bosonic loop by another factor 2 due to the distinction
of particle and anti-particle in case of fermion.  In the presence of
non-vanishing pion field $g^2 \bphi^2$ plays the role of the square of
the effective nucleon mass, as expected simply from the symmetry
consideration.

Adding this term to the effective potential gives rise to new terms in
the stationary conditions for the meson fields and the meson mass
equations.  We could modify our ``Hartree approximation'' by inserting
the following term to the meson self-energy to the right hand side of
(\ref{msigma}) and (\ref{mpi}) :
\begin{equation}
\Pi_{\beta, \sigma}^N ( m_N ) = \Pi_{\beta, \pi}^N ( m_N ) = 
- 8 g^2 \Psi_{\beta} (m_N^2) 
\label{bself}
\end{equation}
with 
\begin{equation}
\Psi_{\beta} ( M ) \equiv \int_F d^4 k \frac{1}{k^2 + M^2} 
\label{bpsi}
\end{equation}
where the factor $8 = 2 \cdot 2 \cdot 2$ in (\ref{bself}) accounts for
the particle-antiparticle, spin, and isospin degeneracy and the
subscript $F$ in (\ref{bpsi}) indicates that the sum is taken over the
fermionic Matsubara frequency.  

This result resembles remarkablly the ones we have obtained for the
meson loop contribution to the meson self-energies (\ref{mself}).
Inclusion of the baryon loop contribution to the meson self-energy is
not a consistent procedure, however, in view of the procedure we have
adopted to drop the diagrams in Fig. \ref{omitted}, since both diagrams
generate non-trivial momentum dependence or dispersion in the
self-energy of mesons and thus again invalidates our ansatz
(\ref{hartreeap}) for the meson propagators.

\begin{figure}
  \epsfxsize= 4 in  \hskip 2 in \special{picture 2}
  \centerline{\epsfbox{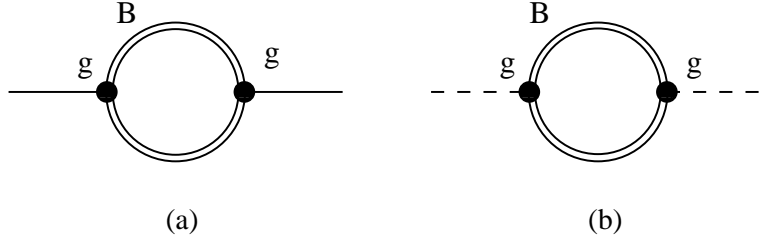}}	  
\vskip 15pt
\caption {Baryon loop diagrams which contribute to the self-energy
of sigma (a) and pion (b)}

\label{baryon-loop}
\end{figure}

\section{High temperature expansion of one-loop integrals}
Phase space integral in one-loop diagrams contain definite integrals
of the following form:
\begin{equation}
I_{\pm}^{(n)} (\mu) \equiv  \int_0^{\infty} \frac{x^n dx}
{\sqrt{x^2 + \mu^2}} \frac{1}{ e^{\sqrt{x^2 + \mu^2}} \pm 1 } .
\end{equation}
For example, the boson (fermion) loop contributions to the meson
self-energies is 
\begin{equation}
\tPhi_\beta (M) = \int \frac{ d^3 k}{(2\pi)^3} \frac{1}{E_k} 
\frac{1}{ e^{\beta E_k} \pm 1 } = \beta^{-2} \frac{1}{2\pi^2}
I_\pm^{(2)} (M\beta), 
\end{equation}
while the pressure of ideal gas of boson (fermion) with mass $M$ 
is 
\begin{equation}
P (\beta; M)  \equiv  
\pm \beta^{-1} \int \frac{d^3 k}{(2\pi)^3}
\ln \left[ 1 \pm e^{-\beta \sqrt{k^2 + M^2}} \right] 
=  \beta^{-4} \frac{1}{6\pi^2} I_\pm^{(4)} ( M \beta ) ,
\end{equation}
where $+(-)$ sign is for fermion (boson).  
  
The functions $I^{(n)}_{\pm} (\mu)$ satisfy the following recursion 
relations:
\begin{equation}
\frac{d}{d \mu^2} I^{(n)}_{\pm} (\mu) = - \frac{n-1}{2}I^{(n-2)}_{\pm}
(\mu)
\label{Idif}
\end{equation}
and have the following limits for $\mu \to 0$:
\begin{eqnarray}
I^{(n)}_- (0) & = & \Gamma (n) \zeta (n), 
\label{Iboundb} \\
I^{(n)}_+ (0) & = & (1-2^{1-n}) \Gamma (n) \zeta (n), 
\label{Iboundf} 
\end{eqnarray}
where $\Gamma (x) $ is the gamma function and $\zeta (x) =
\sum_{m=1}^{\infty} 1/m^x$ is the Riemann $\zeta$ function.  

We wish to evaluate the integral $ I^{(n)}_{\pm} (\mu) $ for the small
nonvanishing value of $\mu = M \beta$, corresponding to high
temperatures ($T >> M$), however, the power series expansion of
$I^{(n)}_{\pm} (\mu)$ would break down at $\mu^n$ since the gamma
function $\Gamma (x)$ is singular at negative integer values of $x$.
These singularities originate from the infrared (small $x$) part of
the integral $I^{(n)}_\pm$. In the case of boson it is enhanced due to
the additional singular behavior of the distribution function of
massless boson at small $x$: $1/(e^x - 1) \sim 1/x$.

Series expansion of $I_{\pm}^{(n)} (\mu)$ was derived by Dolan and
Jackiw in \cite{HighT} for even integer value of $n$.  Here we quote
some of their useful results: 
\begin{eqnarray}
I^{(0)}_- & = & \frac{\pi}{2\mu} + \half \ln \frac{\mu}{4\pi} + \half
\gamma + \half \sum_{n = 1}^\infty \frac{1}{n}
\left[ \left(1+ \frac{\mu^2}{4 \pi^2 n^2} \right)^{-1/2} - 1 \right]
\nonumber \\
& = & \frac{\pi}{2\mu} + \half \ln \frac{\mu}{4\pi} + \half \gamma 
+ \frac{\zeta (3)}{16\pi} \mu^2 - \frac{3 \zeta (5)}{64\pi^2} \mu^4 
+ {\cal O} (\mu^6) ,
\label{I0b} \\
I^{(0)}_+ & = & - \half \ln \frac{\mu}{\pi} - \half \gamma 
+ \half \sum_{n = 1}^\infty \frac{1}{n} 
\left[ \left( 1+ \frac{\mu^2}{4 \pi^2 n^2} \right)^{-1/2} - 1 \right] 
\nonumber \\
& = & - \half \ln \frac{\mu}{\pi} - \half \gamma 
+ \frac{\zeta (3)}{16\pi} \mu^2 - \frac{3 \zeta (5)}{64\pi^2} \mu^4 
+ {\cal O} (\mu^6) 
\label{I0f} 
\end{eqnarray}
where $\gamma = 0.57721 \cdots $ is Euler's number and the numerical
values of the $\zeta$ function at relevant points are $\zeta (2) =
\pi^2/6$, $\zeta (3) = 1.2020 \cdots $, $\zeta (4) = \pi^4/90$, 
$\zeta (5) = 1.0369 \cdots $, and so on.  The series expansion of
$I^{(2)}_{\pm} ( \mu)$ and $I^{(4)}_{\pm} ( \mu)$ can be obtained from
(\ref{I0b}) and (\ref{I0f}) by integrating the differential equation
(\ref{Idif}) with the boundary conditions (\ref{Iboundb}) or
(\ref{Iboundf}): 
\begin{eqnarray}
I^{(2)}_{\pm} (\mu) & = & I^{(2)}_{\pm} (0) - \half 
\int_0^{\mu^2} d\mu'^2 I^{(0)}_{\pm} (\mu'), \nonumber \\ 
I^{(4)}_{\pm} (\mu) & = & I^{(4)}_{\pm} (0) - \frac{3}{2} 
\int_0^{\mu^2} d\mu'^2 I^{(2)}_{\pm} (\mu'). \nonumber 
\end{eqnarray}
The result is 
\begin{eqnarray}
I^{(2)}_- (\mu) & = & \frac{\pi^2}{6} - \frac{\pi }{2 } \mu 
- \frac{1}{4}\mu^2 \ln \frac{\mu}{4\pi} 
+ \left( \frac{1}{8} - \frac{1}{4} \gamma \right) \mu^2 
- \frac{\zeta (3)}{32\pi} \mu^4 + {\cal O} (\mu^6) ,
\label{I2b} \\
I^{(4)}_- (\mu) & = & \frac{\pi^4}{15} - \frac{\pi^2 }{4} \mu^2 
+ \frac{\pi}{2} \mu^3 + \frac{3}{16} \mu^4 \ln \frac{\mu}{4 \pi}
- \frac{3}{8}\left( \frac{3}{4} - \gamma \right) \mu^4  
+ {\cal O} (\mu^6) ,
\label{I4b} 
\end{eqnarray}
for boson and 
\begin{eqnarray}
I^{(2)}_+ (\mu) & = & \frac{\pi^2}{12} + \frac{1}{4} \mu^2 
\ln \frac{\mu}{\pi} + \frac{1}{4} \left( - \half + \gamma \right) \mu^2
- \frac{\zeta (3)}{32\pi} \mu^4 + {\cal O} (\mu^6) ,
\label{I2f} \\
I^{(4)}_+ (\mu) & = & \frac{7\pi^4}{120} - \frac{\pi^2}{8} \mu^2 
- \frac{3}{16} \mu^4 \ln \frac{\mu}{\pi} 
- \frac{3}{32} \left( \frac{3}{2} - 2 \gamma \right) \mu^4
+ {\cal O} (\mu^6) ,
\label{I4f} 
\end{eqnarray}  
for fermion.


\begin{table}[t]
\narrowtext

\begin{tabular}{rrl}
$\lambda$ & $m_\sigma$ (MeV) & ${\teps} = \epsilon / ( m_0^2 \phi_0 )
$ \\
\hline
50&404&0.219 \\ 
100&555&0.0846 \\
200&772&0.0378 \\
500&1208&0.0142 \\ 
1000&1703&0.0069 \\
\end{tabular}
\vskip 1cm
\caption{$\lambda$ dependence of the sigma meson mass at zero
temperature and the scaled symmetry breaking parameter}
\label{table1}   
\end{table}

\end{document}